\documentstyle[12pt]{article}
\title{Rigid Invariance in Gauge Theories}
\begin{document}

\begin{titlepage}
%
\renewcommand{\thefootnote}{\fnsymbol{footnote}}
\begin{flushright}
Februar 1997
\end{flushright}
\vspace{1cm}
 
\begin{center}
{\Large {\bf Rigid Invariance in Gauge Theories}}\footnote{
To appear in  the``Proceedings of 
       the XXIst Conference on Group Theor. Methods in Physics, 
       Goslar, Juli 1996''.} \\[4mm]
{\makebox[1cm]{  }       \\[2cm]
{\bf Elisabeth Kraus}\footnote{supported
 by Deutsche Forschungsgemeinschaft}\\ [3mm]
{\small\sl Physikalisches Institut, Universit\"at Bonn} \\
{\small\sl Nu\ss allee 12, D-53115 Bonn, Germany}} 
{and}\\[0.5cm]
{\bf Klaus Sibold }\\ [3mm]
{\small\sl Institut f\"ur Theoretische Physik, Universit\"at Leipzig} \\
{\small\sl Augustusplatz 10/11, D-04109 Leipzig, Germany} \\[0.5cm]
\vspace{2.0cm}
 
{\bf Abstract}
\end{center}
\begin{quote}
Rigid gauge invariance comprises the symmetry content for physical quantities
in a local gauge theory. Its derivation form BRS invariance is thus crucial for
determining the physical consequences of the symmetry.
\end{quote}
\vfill
\renewcommand{\thefootnote}{\arabic{footnote}}
\setcounter{footnote}{0}
\end{titlepage}

\section{Introduction}
Local gauge invariance and its translation into BRS invariance define gauge models
in the sense that they permit the proof of unitarity. They govern the behaviour
of the unphysical modes which -- in perturbation theory -- have to be introduced 
in order to maintain Lorentz invariance and locality. What they do not tell in 
the general case is which symmetry relations survive quantization and 
renormalization. The existence of conserved currents and charges has to be 
inferred from the rigid gauge invariance or (in the BRS case) from local
Ward identities associated with the rigid symmetry. In this note we shall
sketch some of the more relevant cases.

\section{Quantum electrodynamics}
The classical action
\begin{equation}
\Gamma=\int -\frac 1 4 F_{\mu\nu}F^{\mu\nu}-\frac 1{2\alpha}(\partial A)^2
+\bar{\psi}(i\partial^{\mu}\gamma_{\mu}-m+e A^{\mu}\gamma_{\mu})\psi
\end{equation}
satisfies the local Ward identity (WI)
\begin{equation}
w \Gamma \equiv (-\partial \frac{\partial}{\partial A} -ie\bar{\psi}
\frac{\delta}{\delta\bar{\psi}} +ie\psi\frac{\delta}{\delta\psi}) \Gamma
=-\frac 1 \alpha \Box\partial A\\
\end{equation}
and the rigid Ward identity
\begin{eqnarray}
{\cal W} \Gamma &\equiv& \int \hat{w} \Gamma = 0\\
\hat{w} &\equiv& -ie\bar{\psi}\frac{\delta}{\delta\bar{\psi}}+ie\psi
\frac \delta {\delta\psi}
\end{eqnarray}
Rewriting the local WI for the general Green functions one can prove that
\begin{equation}
\Box \partial A^{Op} = 0,
\end{equation}
i.e. $\partial A ^{Op}$ is a free field operator and can thus be decomposed
into positive and negative frequency part, a decomposition which in turn allows 
to single out physical states by the condition
\begin{eqnarray}
(\partial A) ^{(-)} \mid phys >&=&0.
\end{eqnarray}
This guarantees the absence of negative norm states and leads to the Hilbert 
space of physical states via the formation of equivalence classes (members differ
only by zero norm states) and closure.\\
For the perturbative existence of QED it is now crucial that for every 
regularization or renormalization scheme the local WI  (2) can be proven if 
$e\rightarrow e'=e+o(\hbar), m\rightarrow m'=m+o(\hbar)$ and 
suitable counter terms are added to $\Gamma$. Because then (5) again follows, 
now to all orders of perturbation theory, and (6) together with the construction 
of the state space is possible also.\\
The rigid WI (3) expresses the symmetry content of the theory which just
means conservation of the electric charge. It also expresses the conservation 
of the respective current
\begin{equation}
\hat{w}\Gamma= \partial^{\mu}j_{\mu} 
\end{equation}
whose 0-th component gives rise to the charge once it is integrated over 
3-space. Obviously the WI (3) follows from (2) by integration and (7) differs
from (2) by $-\frac 1 \alpha \Box\partial A +\partial \frac {\delta}
{\delta A}.$ Hence the local WI (2) encodes also all information about the 
symmetry of the theory.\\
This property is a peculiarity of QED which does not hold in models where BRS 
invariance plays a role. These are the models with non-abelian gauge group 
and (or) spontaneous symmetry breaking. 
\section{The abelian Higgs model}
The classical action 
\begin{eqnarray}
\Gamma_{inv}&=&\int(-\frac 1 4 F_{\mu\nu} F^{\mu\nu}+(D_{\mu}\phi)
^*(D^{\mu}\phi)+\frac 1 2 m^2_H\phi^*\phi -\frac 1 2 \frac {m^2_H}
{m^2} e^2(\phi^*\phi)^2)\\
F_{\mu\nu}&\equiv&\partial_{\mu}A_{\nu}-\partial_{\nu}A_{\mu}
\hspace{0,5cm}\phi\equiv\frac 1{\sqrt{2}}(\phi_1+\frac m e +i\phi_2)
					  \hspace{0,5cm}
D_{\mu}\phi\equiv\partial_{\mu}\phi-ie A_{\mu}\phi
\end{eqnarray}
is invariant under the U(1)-transformation
\begin{eqnarray}
\delta_{\omega}\phi&=&ie\omega\phi,\hspace{0,5cm}
\delta_{\omega}A_{\mu}=0.
\end{eqnarray}
It describes a vector field $A_{\mu}$ of mass $m$ in interaction with the 
Higgs field $\phi_1$ of mass $m_H$ and the would-be Goldstone field $\phi_2$.
But
due to the spontaneously broken character of the symmetry the unphysical 
modes of $A_{\mu}$ interact and there is no local gauge WI which would 
characterize the model and guarantee unitarity. It rather has to be replaced
by BRS invariance expressed via a Slavnov--Taylor identity: After introducing 
the Faddeev--Popov ghosts, a Lagrange multiplier field B and suitably coupled 
external fields we arrive at
\begin{eqnarray}
sA_{\mu}&=&\partial_{\mu}c\hspace{1,5cm}s\bar{c} =B\\
\nonumber sc&=&0\hspace{1,8cm} sB =0\\
\nonumber s\phi &=& ie c\phi
\end{eqnarray}
for the elementary fields in the tree approximation and at
\begin{equation}
s(\Gamma)\equiv \int\partial_{\mu}c \frac{\delta\Gamma}{\delta A_{\mu}}+B\frac
{\delta\Gamma}{\delta\bar{c}}+\frac{\delta\Gamma}{\delta Y_{i}}\frac
{\delta\Gamma}{\delta\phi_{i}}=0.
\end{equation}
For calculational purposes one prefers a 't Hooft type gauge fixing
\begin{equation}
\frac {\delta\Gamma}{\delta B}= \xi B+\partial A +\xi_{A}m\phi_2
\end{equation}
which however breaks the naive rigid invariance as defined by (10). The 
construction of higher orders thus requires some more machinery (i.e. further 
external fields) and in particular leads to a deformation of the classical
rigid invariance in a well-specified sense.
I.e. if one imposes physical normalization
conditions for the vector and Higgs mass and their wave function
renormalizations, 
then all normalizations of the wave function of $\phi_2$ at finite
momentum lead  to a rigid WI of the type
\begin{equation}
{\cal W} \Gamma\equiv\int(-\phi_2\frac{\delta}{\delta\phi_1}+(1+u)(\phi_1-\hat{\xi}
_A\frac m e)\frac{\delta}{\delta\phi_2})\Gamma=0
\end{equation}
(for all external fields = 0). $u=o(\hbar)$ parametrizes the deformation.
It is crucial for the derivation of (14) that the rigid WI operator ${\cal W}$
is symmetric with respect to the ST identity (12). The coupling has been
fixed by 
some 3-point-function. It is clear that to (14) is associated a local WI.
Although
it does not define the model and does not yield unitarity it is nevertheless
useful. One can show [1] that it has the form
\begin{equation}
(e(1+a) w (x)-\partial\frac{\delta}{\delta A})\Gamma = \Box B
\end{equation}
(Here $ \int w(x) =  {\cal W}$.)
If one refines the argument by varying the gauge parameter $\xi$ into a
Grassmann variable $\chi$, adds this variation to the ST identity and ensures
this enlarged identity to all orders one has control over the gauge parameter
dependence. It turns out [2] that the coefficient $a$ in (15) is independent
of the gauge parameter. Hence one can like in QED {\it define} the coupling via the 
requirement ``validity of an exact local WI to all orders''
\begin{equation}
(ew(x)-\partial\frac{\delta}{\delta A})\Gamma=\Box B.
\end{equation}
A conserved current is defined via (16) by
\begin{equation}
ew(x)\Gamma =\partial^{\mu}j_{\mu}\cdot\Gamma
\end{equation}
The associated charge does, however, not exist since the symmetry is 
spontaneously broken.
\section{QCD}
As an example of a Yang-Mills theory with unbroken gauge group we look at
QCD i.e. $SU(3)$ and multiplets of fermions in the fundamental representation.
The classical invariant action has the form
\begin{eqnarray}
\Gamma_{inv}&=&\int-\frac 1 {4 g^2}Tr F_{\mu\nu}F^{\mu\nu}+\bar{\psi}
(iD^{\mu}\gamma_{\mu}-m)\psi\\
F_{\mu\nu}&\equiv&(\partial_{\mu}A_{\nu}^a-\partial_{\nu}A_{\mu}^a+f^{abc}
A_{\mu}^bA_{\nu}^c)\tau^a\\
\nonumber D_{\mu}\psi&\equiv& (\partial_{\mu}-ie\tau^a A_{\mu}^a)\psi\\
\nonumber[\tau^a,\tau^b]&=&if^{abc}\tau^c
\end{eqnarray}
Since classically rigid invariance
\begin{equation}
{\cal W}^a\Gamma_{inv}\equiv i\int(A_{\mu}^bf^{bac}\frac
{\delta}{\delta A_{\mu}^c}-\bar{\psi}i\tau^a\frac{\delta}{\delta\bar{\psi}}
+\psi i\tau^a\frac{\delta}{\delta\psi})\Gamma=0
\end{equation}
is not broken one tries to impose (20) to all orders. A theorem due to 
BRS [3,4] guarantees that this is for (semi-) simple gauge groups indeed
possible (in any renormalization scheme). In a second step one establishes 
then BRS invariance by way of a ST identity: 
\begin{equation}
s(\Gamma)\equiv\int(\frac{\delta\Gamma}{\delta\varrho}\frac{\delta\Gamma}
{\delta A}
+B \frac {\delta\Gamma}{\delta\bar{c} }+\frac{\delta\Gamma}{\delta\sigma}\frac
{\delta\Gamma}{\delta c}+\frac{\delta\Gamma}{\delta\bar{Y}}\frac
{\delta\Gamma}{\delta\psi}+\frac{\delta\Gamma}{\delta\bar{\psi}}
\frac{\delta\Gamma}{\delta Y}=0
\end{equation}
The existence of a local WI and its relation to a conserved current becomes a 
rather subtle question because the $\phi\Pi$ ghosts do not couple minimally
to the vector field. A suitable device for their construction is an external 
field $\tilde{A}_{\mu}$, called background field, which varies under BRS in 
a further Grassmann field $\tilde{c}_{\mu}$ [5] :
\begin{equation}
s\tilde{A}_{\mu}=\tilde{c}_{\mu}
\end{equation}
(All of them transform according to the adjoint representation under rigid 
symmetry.) Taking also gauge parameter variation into account one can
derive [6]
\begin{equation}
(gw^a-\partial\frac{\delta}{\delta A^a}-\partial \frac{\delta}
{\delta\tilde{A}^a})\Gamma = 0,
\end{equation}
\begin{equation}
w^a\equiv \int\sum_{\phi} f^{bac}\phi^b\frac{\delta}{\delta\phi^c}
+i\bar{\psi}\tau^a\frac{\delta}{\delta\bar{\psi}}-i\psi\tau^a\frac{\delta}
{\delta\psi}\hspace{0,5cm}\phi:A,\tilde{A},B,c,\bar{c},\tilde{c}_{\mu}
\end{equation}
Here, remarkably enough, one can again define the coupling g by the requirement
``validity of exact local WI''. (``Exact'' meaning that it holds in this form 
without quantum corrections.)
\section{The electroweak standard model}
The symmetry group of the electroweak standard model (SM) is
$SU(2)\times U(1)$, hence not
semi-simple. It is furthermore spontaneously broken bringing about the 
difficulty of identifying an unbroken U(1) subgroup which eventually yields the
electric charge. The most systematic way of proceeding is to impose
and establish
first of all the BRS invariance associated with the gauge group, hence to
forget 
about rigid invariance and choosing a gauge fixing and $\phi\Pi$-sector
so general,
that one can encompass all possible situations [7]. In order to analyze
the possible
forms of rigid transformations in higher orders one writes down a set of WI
operators
${\cal W}^a\hspace{0,5cm}(a =+,-,Z,A)$ constrained only by formal charge
conservation and by
commutation with the ST identity. Then one requires an $SU(2)\times U(1)$
algebra for them
and solves for the most general representation matrices in all relevant
field sectors
(vector, scalar, spinor). Under the simplifying assumption of CP-invarince (no 
family mixing) we have identified all free parameters appearing in ${\cal W}^a$
with free one's appearing in the general solution of the ST identity [8]. Like
in the abelian Higgs model it is then clear how normalization conditions
lead to
deformed rigid WI and the associated deformed algebra.

\section*{Acknowledgments}
K.S. thanks the organizers for the invitation to this nice conference.\\

\end{document}